\documentclass[twocolumn,aps,showpacs,prb,superscriptaddress]{revtex4-1}
\usepackage{amsmath,amssymb}
\usepackage{subfig}
\usepackage{graphicx}
\usepackage{braket}
\usepackage{epstopdf}
\usepackage[usenames]{color}
\usepackage{hyperref}
\usepackage{float}
\usepackage{xcolor}
\usepackage{tikz}
\usetikzlibrary{positioning}

\begin{document}

\title{Many Body Localization Due to Correlated Disorder in Fock Space}

\author{Soumi Ghosh}
\affiliation{Department of Physics, Indian Institute of Science, Bangalore 560 012, India}
\author{Atithi Acharya}
\affiliation{Department of Physics, Indian Institute of Science, Bangalore 560 012, India}
\affiliation{Department of Physics and Astronomy, Rutgers University, NJ 08854, USA}
\author{Subhayan Sahu}
\affiliation{Department of Physics, Indian Institute of Science, Bangalore 560 012, India}
\affiliation{Condensed Matter Theory Center and Department of Physics, University of Maryland, College Park, MD 20742, USA}
\author{Subroto Mukerjee }
\affiliation{Department of Physics, Indian Institute of Science, Bangalore 560 012, India}

\begin{abstract}
In presence of strong enough disorder one dimensional systems of interacting spinless fermions at non-zero filling factor are known to be in a many body localized phase.
When represented in `Fock space', the Hamiltonian of such a system looks like that of a single `particle' hopping on a Fock lattice in the presence of a random disordered potential. The coordination number of the Fock lattice increases linearly with the system size $L$ in one dimension. Thus in the thermodynamic limit $L\rightarrow \infty$, the disordered interacting problem in one dimension maps on to an Anderson model with infinite coordination number. Despite this, this system displays localization which appears counterintuitive. A close observation of the on-site disorder potentials on the Fock lattice reveals a large degree of correlation among them as they are derived from an exponentially smaller number of on-site disorder potentials in real space. This indicates that the correlations between the on-site disorder potentials on a Fock lattice has a strong effect on the localization properties of the corresponding many-body system. This intuition is also consistent with studies of quantum random energy model where the typical mid-spectrum states are ergodic and the on-site potentials in Fock space are completely uncorrelated. In this work we perform a systematic quantitative exploration of the nature of correlations of the Fock space potential required for localization. We study different functional variations of the disorder correlation in Fock lattice by analyzing the eigenspectrum obtained through exact diagonalization. Without changing the typical strength of the on-site disorder potential in Fock lattice we show that changing the correlation strength can induce thermalization or localization in systems. From among the various forms of correlations we study, we find that only the linear variation of correlations with Hamming distance in Fock space is able to drive a thermal-MBL phase transition where the transition is driven by the correlation strength. Systems with the other forms of correlations we study are found to be ergodic.
\end{abstract}


\maketitle
\section{Introduction}
It was argued by Anderson in a seminal paper~\cite{Anderson.1958} that for a system of non-interacting particles, sufficiently strong disorder can localize all energy eigenstates. In dimensions $d\le2$ any arbitrarily weak amount of disorder can induce localization of all eigenstates~\cite{TVR.1979,TVR.1985}.
A generalization of this phenomenon in the presence of interaction, known as Many Body Localization(MBL)\cite{BASKO.2006} has been shown to exist (at least in one dimension) and has attracted a lot of interest recently~\cite{Abanin.2013,Huse.Pheno.2014,Lfconductivity.2015,GE.MBL.2015,QPMBL.2013,SPME.Modak,Non-ergodic-metal,Huse.Universality,Potter.2017}.
As a consequence of localization and hence lack of diffusion, isolated MBL systems fail to thermalize on their own\cite{PRB.Huse.2007,PalHuse.2010} and thus do not obey the Eigenstate Thermalization Hypothesis(ETH)\cite{QSM.closed.1991,Srednicki.1994,Rigol2008}.
Thus systems exhibiting MBL can retain their memory of initial conditions for arbitrarily large times despite the presence of interactions which cause some amount of dephasing.
It has been argued that this non-ergodic behavior of MBL systems can also be understood in terms of emergent conservation laws which dynamically prevent thermalization~\cite{modak1.2015,Huse.Pheno.2014,Serbyn.2013,Chandran.2015} similar to the mechanism of non-ergodicity in traditional integrable systems~\cite{Integrable.Rigol,Chaos.Rigol,Integrable.Modak,RME.Modak}.
Typically, for small enough disorder strength these systems display ergodicity and undergo a thermal-
MBL transition with increasing disorder strength~\cite{PRB.Huse.2007,PalHuse.2010}.
The thermal-MBL transition is not a regular phase as encountered in equilibrium statistical mechanics and instead is a dynamical phase transition where the nature of the dynamics of the system, as encoded in its eigenspectrum, changes from being ergodic to non-ergodic. A signature of this transition is the entanglement properties of the eigenstates which go from being volume law entangled with thermal values of the entanglement entropy (on the ergodic side) to being area law entangled with non-thermal values of the entropy (on the MBL side).

One of the central models for studying MBL is a one dimensional system of interacting spinless fermions with disorder at non-zero filling factors~\cite{PRB.Huse.2007}, which we refer to henceforth as the interacting Anderson model. The Hamiltonian for this model is 
\begin{equation}
H = \sum_{\langle i j \rangle} t_{ij} c_i^\dagger c_j + {\rm h.c.} + \sum_i \epsilon_i n_i + \sum_{ij} V_{ij} n_i n_j,
\label{Eqstdham} 
\end{equation}
where $\langle i j \rangle$ label pairs of sites with hopping $t_{ij}$ and interaction $V_{ij}$ between them. $\epsilon_i$ labels the on-site disorder. An alternative basis in which this Hamiltonian can be written is the so-called `Fock basis'. The basis states are specified by the occupancies of each site $| \{ n_i\} \rangle$ consistent with the filling factor. The above Hamiltonian in this representation can be written as 
\begin{eqnarray}\nonumber
H &=& \sum_{\{ n_i \}, \{ m_i \}} J_{\{ n_i \}, \{ m_i \}} | \{ n_i\} \rangle \langle \{ m_i\} | + {\rm h.c.}\\
& & + \sum_{\{ n_i\}} U_{\{ n_i\}} | \{ n_i\} \rangle \langle \{ n_i\} |
\label{Eqfockham}
\end{eqnarray}
which has the form of a `tight-binding' model in Fock space with the `hopping' matrix $J_{\{ n_i \}, \{ m_i \}}$ being a function of the real space hopping amplitudes $t_{ij}$ in Eqn.~\ref{Eqstdham} and the `on-site potential' $U_{\{ n_i\}}$ depending on the real space
potentials $\epsilon_i$ and density-density interaction $V_{ij}$. One can also define a notion of distance on the Fock lattice in terms of the Hamming distance employed in information theory. The Hamming distance between two states $| \{ n_i\} \rangle$  and $| \{ m_i\} \rangle$ is the number of sites which have different occupancies in the two states. One can cast any Hamiltonian of the form of Eqn.~\ref{Eqstdham} (even when the $\epsilon_i$'s are not random variables) in the form of the tight-binding Hamiltonian in Eqn.~\ref{Eqfockham} in Fock space. In particular, setting $V_{ij}=0$ in Eqn.~\ref{Eqstdham} and letting $\epsilon_i$ correspond to uncorrelated disorder gives the Hamiltonian for Anderson localization in one dimension, where it is known that all single particle eigenstates are localized when $t_{ij}$ is not long ranged. All many-body eigenstates of this system are trivially many-body localized since they are obtained by occupying localized single particle eigenstates. 

One can also study the above problem of Anderson localization in Fock space. The hopping term $J_{\{ n_i \}, \{ m_i \}}$ on the Fock lattice for any $t_{ij}$ is zero unless the Hamming distance between $| \{ n_i\} \rangle$ and $| \{ m_i\} \rangle$ is exactly equal to two. The number of such sites for a given site $| \{ n_i\} \rangle$ can be thought of as the co-ordination number (i.e. the number of sites accessible in one hop) on the Fock lattice. It is straightforward to see that the co-ordination number on the Fock lattice increases linearly with the size of the real space lattice $L$ at fixed filling. Thus, in the thermodynamic limit $L \rightarrow \infty$, the co-ordination number tends to $\infty$ as well. The on-site potentials $U_{\{ n_i\}}$ are linear combinations of the random disorder potentials $\epsilon_i$ and hence random themselves. Thus, it might appear that the Anderson problem in the thermodynamic limit in real space can be mapped on to an Anderson problem on the Fock lattice with infinite co-ordination number. However, this presents a paradox since the Anderson model on a lattice with infinite co-ordination number is not expected to have any localized states~\cite{e-expansion}. The resolution lies in the fact that the potentials $U_{\{ n_i\}}$, while random, are not uncorrelated since there are $\sim e^L$ of them which are derived from only $L$ random variables $\epsilon_i$. Thus, the equivalent Anderson problem on the Fock lattice involves correlated on-site potentials and it has been shown using field theoretic methods that such correlations can indeed lead to localization in lattice with infinite co-ordination number~\cite{PRL.Altland}. In this context, it is interesting to note that the random energy model~\cite{REM.Derrida}, which has been extensively studied in the context of spin glasses, has energy eigenstates with MBL when subjected to quantum fluctuations introduced through a transverse field~\cite{PRB.Laumann,PRL.Laumann}. However, the relevant model called the Quantum Random Energy Model (QREM), has states with MBL only near the edges of its energy spectrum with the mid-spectrum states remaining ergodic. When the QREM is cast in the form of Eqn.~\ref{Eqfockham}, the on-site potentials $U_{\{ n_i\}}$ depend not just on $L$ random variables of the form of $\epsilon_i$ but a larger ($\sim e^L$) number of random variables corresponding to different $N$-body random interactions. Thus, the $U_{\{ n_i\}}$ are no longer strongly correlated with one another resulting in the delocalization (ergodicity) of typical mid-spectrum states, consistent with the notion of localization being induced by the correlation of potentials in Fock space. In another relevant model it was shown that in the presence of correlations there could be clusters of energy eigenstates which display level repulsion within regions of Fock space but no spectral correlations across the entire space\cite{Laumann-corr}. Such states look indistinguishable from localized states when level statistics is employed as a diagnostic. Very recently, features related to MBL and the MBL transition in Hamiltonians of the form in Eqn.~\ref{Eqfockham} have been examined through calculations of Greens functions and self energies in the thermodynamic limit~\cite{Logan.Fockspace1,Logan.Fockspace2}. These studies also highlight the importance of the correlations of on-site potentials. 

An important issue that has so far not received much attention and which we address in this paper is a systematic quantitative exploration of the nature of correlations of the Fock space potential required for localization.  We use the covariance between 
potentials, treated as random variables, as a measure of the correlation among them. The covariance is expressed as a function of the Hamming distance between basis states and we consider several different functional forms and we perform numerical exact diagonalization to obtain the eigenspectrum. Our main result is that from among several natural functional forms of the covariance, only the one with a linear variation of the covariance with Hamming distance displays an ergodic-MBL transition as a function of correlation strength while the others seem to always yield ergodicity. The rest of the paper is organized as follows: In section II we describe the general form of Fock space Hamiltonian used for our studies. We also explicitly obtain the correlations among the on-site Fock space potentials as a function of Hamming distance for the 1D Anderson Hamiltonian and describe other natural functional forms of the correlations that we study. In section III, we outline the method to generate correlated random numbers and also described the analysis performed to identify the ergodic and MBL phases and the transition between them. Finally, in section IV we present our results and discuss how the correlation among on-site terms in Fock space affects the ergodicity or localization of the eigenstates of the corresponding Hamiltonians.

\section{Model}
We study one dimensional systems of $N$ sites and $m$ spinless fermions described by the Hamiltonian in Eqn.~\ref{Eqfockham}. Thus, the Fock space has $^NC_m$ basis states each denoted by a set of $n_i$'s where $n_i$ is $1$($0$) if the site $i$ in real space is occupied(unoccupied). Therefore there are exactly $m$ values of $n_i$'s which are equal to $1$ and $(N-m)$ which are equal to zero for each basis state and each such state is a different combination of the $1$'s and $0$'s. The basis states correspond to vertices (sites) of a hypersolid in Fock space. The Hamiltonian of Eqn.~\ref{Eqfockham} defined on these sites is a sum of two terms
\begin{eqnarray} \nonumber
& &H = H_{hop}+H_{onsite},\\ \nonumber
& &H_{hop} = \sum_{\{n_i\},\{m_i\}} J_{\{n_i\},\{m_i\}}\ket{\{n_i\}}\bra{\{m_i\}}+ h.c.\\ 
& &H_{onsite} = \sum_{\{n_i\}}U_{\{n_i\}}\ket{\{n_i\}}\bra{\{n_i\}}.
\label{eq:ham1}
\end{eqnarray}
$U_{\{n_i\}}$ is the potential at site $\{n_i\}$ and when Eqn.~\ref{eq:ham1} is derived from a real space Hamiltonian depends on the the density dependent terms (on-site potentials and interactions) of that Hamiltonian. 
For such a case the hopping terms $J_{\{n_i\},\{m_i\}}$ between the Fock space sites $\{n_i\}$ and $\{m_i\}$ are determined by the hopping amplitudes in real space and are given by:
\begin{eqnarray}\nonumber
J_{\{n_i\},\{m_i\}}&=&\bra{\{n_i\}}\sum_{i}\left( -t_1 c^\dagger_{i+1}c_i-t_2 c^\dagger_{i+2}c_i-\dots\right.\\
& &\left.-t_p c^\dagger_{i+p}c_i+h.c.\right)\ket{\{m_i\}}
\label{eq:hop}
\end{eqnarray}
where $t_1$,$t_2$,$t_p$ are respectively nearest neighbor, next nearest neighbor and $p^{th}$ neighbor hopping amplitudes in real space. As each of these terms involves the hop of a single fermion in real space,  two Fock space sites can have a non-zero hopping integral $J_{}$ only when the hamming distance between them is equal to two. 

Now let us consider the interacting Anderson model with nearest neighbor and next nearest neighbor hopping
\begin{eqnarray} \nonumber
H&=&\sum_i \bigl[ \varepsilon_i n_i-t_1(c_i^\dagger c_{i+1}+h.c.)-t_2(c_i^\dagger c_{i+2}+h.c.)\bigr.\\
& &\bigl. +Vn_in_{i+1}\bigr]
\label{eq:anderson}
\end{eqnarray}
where the $\varepsilon_i$'s are random on-site potentials in real space, $V$ is the nearest neighbor interaction and $t_1$ and $t_2$ are nearest neighbor and next nearest neighbor hopping respectively. This Hamiltonian written in Fock space has the form of  Eqn.~\ref{eq:ham1} with $H_{onsite}$ given by the on-site disorder and interaction terms and $H_{hop}$ given by Eqn.~\ref{eq:hop} with $t_p=0$ for $p>2$.
The on-site term $U_{\{n_i\}}$ at each Fock space site is solely determined by the occupation numbers $n_i$'s of the corresponding real space sites labeled by the $i$'s.
\begin{equation}
U_{\{n_i\}}=\sum_i \varepsilon_i n_i+Vn_in_{i+1}
\label{eq:onsite}
\end{equation}
Evidently, the $U_{\{n_i\}}$'s are random due to the randomness of the $\varepsilon_i$s. However, they are correlated as discussed earlier~\cite{PRL.Altland} because for each disorder realization there are $N$ random on-site energies $\varepsilon_i$ in real space out of which $m$ are summed up in Eqn.~\ref{eq:onsite} depending on the occupation set $\{n_i\}$. Thus $^NC_m$ ($\sim e^N$ for large $N$ and fixed filling $m/N$) random $U_{\{n_i\}}$'s are generated from $N$ number of random $\varepsilon_i$'s which gives rise to a large degree of correlation between the Fock space on-site potentials.

Qualitatively it can be seen that if two Fock space sites $\{n_i\}$ and $\{m_i\}$ have $n_i=m_i$ for most of the sites $i$, then the on-site terms corresponding to these two Fock space sites will be more correlated than when for most of the real space site indices $i$, $\{n_i\}$ are not equal to $\{m_i\}$. Therefore, the Hamming distance $r$ which is defined by the number of real space sites $i$ for which $n_i \ne m_i$ can be used to quantify the correlation between $U_{\{n_i\}}$ and $U_{\{m_i\}}$. We show (in Appendix-A) that the covariance between the on-site terms in Fock space in this case varies with the Hamming distance $r$ as:
\begin{equation}
Cov(r)=m \sigma^2 \left(1-\frac{r}{N}\right)
\end{equation}
where the $\varepsilon_i$'s are identically distributed random variables with mean zero and standard deviation $\sigma$. We choose $\sigma=1$ for most of our calculations. Note that the variance of $U_{\{n_i\}}$ at every site $\{n_i\}$ in Fock space is $m$. 

In this paper we also study models with other functional forms of the covariance as a function of Hamming distance $r$, not necessarily derivable from simple real space one dimensional models.  In doing so we keep the variance of the on-site terms in Fock space the same as above (i.e. $=m$)\footnote{Here the variance of the on-site terms in Fock space is defined by summing over a large number of different disorder realizations}. This enables us to study only the effect of different functional forms of the covariance on localization without changing the typical strength of the disorder in Fock space. For consistency, the hopping term $H_{hop}$ for each model we study is of the form described in Eqn.~\ref{eq:hop} with $t_p=0$ for $p>2$. The amplitudes of the hopping are chosen such that the disorder strength in Eqn.~\ref{eq:anderson} (which was chosen to be $1$) is strong enough to show MBL in the presence of the interaction $V$ (Eqn.~\ref{eq:anderson}), which is taken to be equal in magnitude to the nearest neighbor hopping $t_1$. Thus, the hopping parameters $t_1$ and $t_2$ are chosen to be $0.2$ and $0.1$. The different forms of covariance considered are listed below:\\
\textbullet \textbf{Case-I:} $Cov(r)=m\delta_{r,0}$, i.e. the on-site terms in Fock space are independent random variables having a variance $m$, where $m$ is the number of particles.\\
\textbullet \textbf{Case-II:} $Cov(r)=m\left(1- \gamma \frac{r}{N}\right)^p$, where $m$ and $N$ are the total number of particles and total number of sites respectively and $\gamma$ and $p$ are adjustable parameters.\\
\textbullet \textbf{Case-III:} $Cov(r)=m\left[(1-a)\left(1-\frac{r}{N}\right)+a\left(1-\frac{r}{N}\right)^2\right]$, where $m$ and $N$ are the same as above and $a$ is the varying parameter. \\
\textbullet \textbf{Case-IV:} $Cov(r)=m\exp \left[-a\frac{r}{N}\right]$, where $m$ and $N$ are the same as before and $a$ is the varying parameter.\\
\textbullet \textbf{Case-V:} $Cov(r)=m\left[u \delta_{r,0}+v\left(1-\delta_{r,0}\right)\right]$, where $u$ and $v$ are the varying parameters. $mu$ is the variance of the disordered on-site terms and $mv$ is the covariance between any two different Fock space sites $\{n_i\}$ and $\{m_i\}$ independent of the Hamming distance between them.

\section{Method and Analysis:}
We generate correlated random numbers with the desired covariance from completely uncorrelated normally distributed random numbers.
For this we first write the covariance matrix in Fock space by calculating the Hamming distances between different pairs of Fock space sites and associating a number to that pair using the chosen functional forms of the covariances. Thus, if a pair of Fock space sites $I$ and $J$ are $r_{IJ}$ hamming distance away, the corresponding matrix element of the covariance matrix becomes $C_{IJ}=C_{JI}=Cov(r_{IJ})$. The covariance matrix, once defined, can be decomposed into two parts $L$ and $L^T$, where
\begin{equation}
\label{eq:decomposition}
C=LL^T
\end{equation}
Now if $X=(x_1; x_2; \dots ;x_f)$ (where f is the Fock space dimension) are uncorrelated normally distributed random numbers with variance $1$, the correlated random numbers $Y=(y_1;y_2;\dots ;y_f)$ are given by:
\begin{equation}
Y=LX
\end{equation}

This can be seen as follows: If $y_I$ and $y_J$ are two on-site terms in Fock space then
\begin{eqnarray}
\braket{y_J y_I}&=&\braket{L_{JK} x_K L_{IK'}x_{K'}}\\ \nonumber
&=&L_{JK}L_{IK'}\braket{x_Kx_{K'}}=L_{JK}L_{IK}=C_{JI}
\end{eqnarray}
where the average $\braket{\dots}$ is performed over different realizations of the random numbers X and we know $\braket{x_Kx_{K'}}=\delta_{K,K'}$,for $x_K$,$x_{K'}$ are uncorrelated random numbers with variance 1.
Note that to have a decomposition like Eqn.~\ref{eq:decomposition} the covariance matrix must be positive semidefinite. However, this does not restrict the study. Rather, the covariance matrix of any random data must be positive semidefinite. Since, the correlated random numbers here are produced from the covariance matrix, a condition of positive semi-definiteness has to be imposed on the chosen forms of covariance matrices.
Once the values of the random potential for the various Fock space sites are generated, we write the Hamiltonian in Fock space(Eqn.~\ref{eq:ham1}) and perform numerical exact diagonalization for system sizes $N=10,12,14,16$.

We analyze the ergodic-MBL transition using the statistics of adjacent energy gaps of the many body Hamiltonian.
It is known that for many body localized states, two eigenstates having similar energies are far apart (where the notion of distance is the Hamming distance) in Fock space  and thus experience no level repulsion.
As a result, successive energy gaps are Poisson distributed~\cite{PRB.Huse.2007}.
On the other hand, in the ergodic phase, the level spacing statistics is that of a Gaussian orthogonal ensemble(GOE).
We calculate the average of the ratio of successive energy gaps $r_n=\frac{min\{\delta_n,\delta_{n+1}\}}{max\{\delta_n,\delta_{n+1}\}}$ where $\delta_n=E_{n+1}-E_n$ is the energy gap between $n^{th}$ and $(n+1)^{th}$ energy level. It is known that the average value of this ratio is $\sim0.529$ for GOE while for a Poissonian distribution it is $\sim0.386$.
Thus, the average level spacing ratio can be used to locate the ergodic-MBL transition as a function of a tuning parameter.
While calculating the level spacing ratio, we average over $5000,2000,500$ and $50$ realizations of the random on-site terms in Fock space for the system sizes $N=10,12,14$ and $16$ respectively.

\section{Results}
In this section we present the results of our studies on the models described earlier and discuss the presence or absence of an ergodic-MBL transition in them.
\paragraph*{\textbf{\emph{Case-I:}}}
\begin{figure}[t]
\centering
\begin{tikzpicture}
\node(img) {\includegraphics[width=\linewidth]{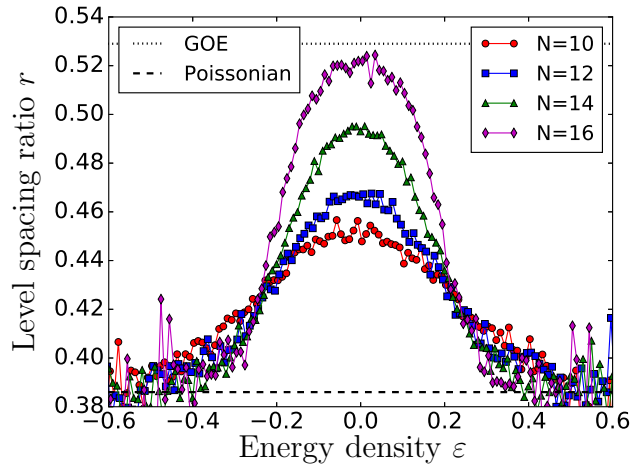}};
\node[below=of img,node distance=0cm,yshift=1.5cm]{\large{Energy density $\varepsilon$}};
\node[left=of img,node distance=0cm,rotate=90,anchor=center,yshift=-1.1cm]{\large{Level spacing ratio $r$}};
\end{tikzpicture}
\caption{Variation of level spacing ratio as a function of the energy density for on-site terms in Fock space with covariance $Cov(r)=m\delta_{r,0}$(Case-I): The states at the middle of the spectrum following GOE statistics are ergodic while the states at the tail of the spectrum following Poissonian statistics are many body localized. There is a many body mobility edge at $E_c \approx \pm 0.2$ separating these two type of states.}
\label{fig:rem}
\end{figure}
The $U_{\{n_i\}}$'s are uncorrelated random numbers with standard deviation $m$. The particular distribution we choose for them is
\begin{equation}
P(U)=\dfrac{1}{\sqrt{\pi N}} e^{-\frac{U^2}{N}}.
\end{equation}
The corresponding model is reminiscent of the quantum random energy model discussed in \cite{PRB.Laumann,PRL.Laumann} but with a difference: There is no transverse field term which induces nearest neighbor hopping between the Fock sites. Instead, the hopping in Fock space is determined by nearest neighbor and next nearest neighbor hopping in real space as described in Eqn.~\ref{eq:hop}. Since we are working at half filling ($m=N/2$), $Cov(r)=m\delta_{r,0}$ for this model.

\begin{figure}[t]
\centering
\begin{tikzpicture}
\node (img) {\includegraphics[width=\linewidth]{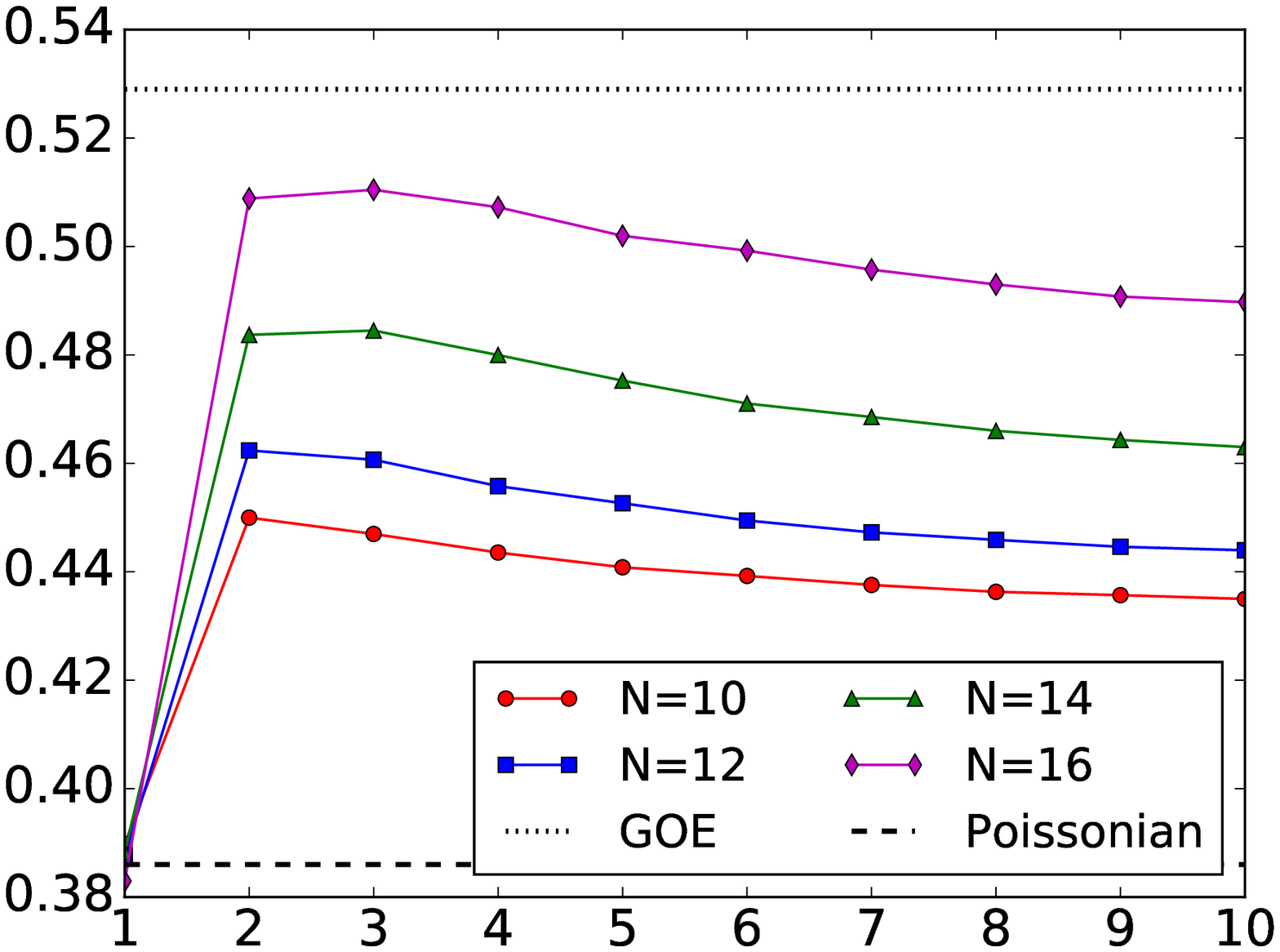}};
\node[below=of img,node distance=0cm,yshift=1.5cm]{\large{$p$}};
\node[left=of img,node distance=0cm,rotate=90,anchor=center,yshift=-1.1cm]{\large{Average level spacing ratio $\braket{r}$}};
\end{tikzpicture}
\caption{Variation of average level spacing ratio with changing exponent $p$ for $Cov(r)=m(1-\frac{r}{N})^p$: Only the $p=1$ case corresponds to the  Poissonian value while for all $p>1$ the average level spacing ratio has a value close to that for GOE. $p=2$ has the maximum value for $\braket{r}$ and it then decreases with increasing $p$ and saturates to a value depending on the system size. The average level spacing ratio is calculated by averaging over $5000,2000,500,50$ samples for $L=10,12,14,16$ respectively.}
\label{fig:r_exp}
\end{figure}

Fig.~\ref{fig:rem} shows the variation of the level spacing ratio as a function of the energy density of the many body eigenstates. The finite-size calculations show that the states with energy density near zero (i.e in the middle of the spectrum) possess Wigner-Dyson (GOE) statistics implying ergodicity while the states with energy densities at the tail of the band possess Poissonian statistics indicating MBL. The many body mobility edge separating these two phases is also evident from the crossover region of these finite size calculations. This result is similar to that of quantum random energy model implying that the different kind of hopping in Fock space influences the spectrum of the many body system very little. On the other hand, the uncorrelated nature of the random disorder in Fock space dominates the behavior of the system.
\paragraph*{\textbf{\emph{Case-II:}}}
Here we calculate the average level spacing ratio for different ranges of the two parameters $\gamma$ and $p$ with a view to locate an ergodic-MBL transition, if one exists. The covariance is 
\begin{equation}
cov(r)=m\left(1-\gamma\frac{r}{N}\right)^p
\label{eq:case2}
\end{equation}
As in the previous case, we also calculate the energy resolved level spacing ratio to locate the presence or absence of a mobility edge in the many body spectrum. We consider specific values of $\gamma$ and $p$ below.
\paragraph{\boldmath{$\gamma=1$}\textbf{\emph{, integer}} $p:$}
Fig.~\ref{fig:r_exp} shows the variation of the average level spacing ratio as a function of the exponent $p$ for $\gamma=1$. Here we choose only integer $p$ as non-integer values of $p$ do not yield positive semidefinite covariance matrices.
$p=1$ corresponds to the interacting Anderson model (Eqn.~\ref{eq:anderson}).

Fig.~\ref{fig:r_exp} shows that only the $p=1$ has the Poissonian value for the average level spacing ratio and for all $p>1$, the average value of the  level spacing ratio is near the GOE value. For $p=2$ the level spacing ratio average $\braket{r}$ is a maximum and decreases with increasing $p$ before saturating at values corresponding to case-I for different system sizes. This implies that for larger values of the exponents $p$, the level spacing statistics approach that for uncorrelated random disorder in Fock space (case-I). This can be understood from the fact that the covariance relation in Eqn.~\ref{eq:case2} falls faster with the hamming distance $r$ as exponent $p$ is increased and for sufficiently large values of $p$ the covariance falls off to zero before the hamming distance reaches 2 (which is the minimum value of the hamming distance possible in our system due to the fact that the number of particles is fixed), which is nothing but the case-I. Although the fine tuning of the parameter $p$ is restricted due to the breaking of positive semi-definiteness of the covariance matrix at non-integer $p$'s, we do not see any crossing between curves for different system sizes implying the absence of a scale invariant critical point for this MBL to thermal transition. We further investigate this in Case-III where we interpolate between $p=1$ and $p=2$ conserving the constraint of positive semi-definite covariance matrix.
\begin{figure}[t]
\begin{tikzpicture}
\node (img1) {\includegraphics[width=\linewidth]{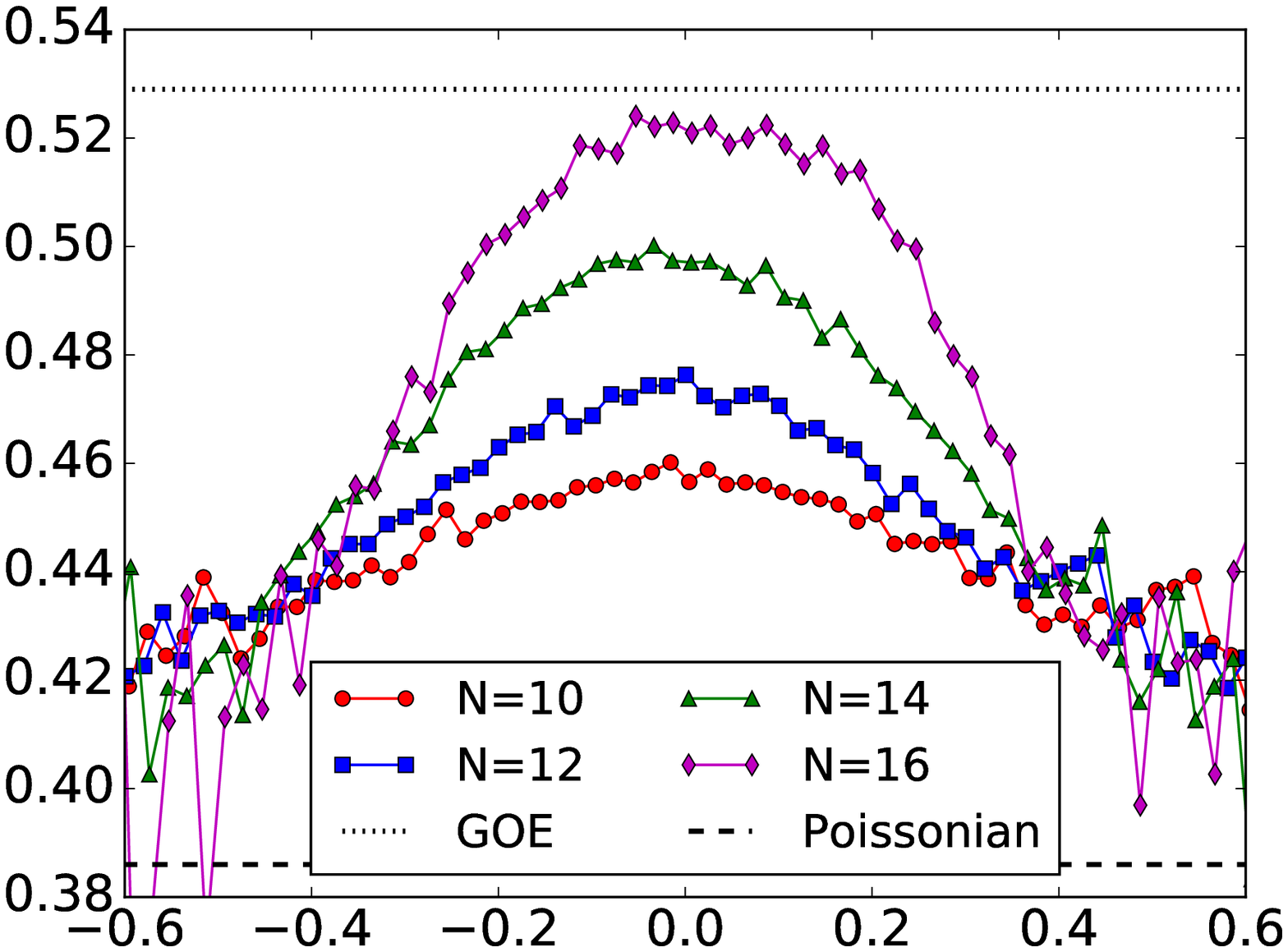}};
\node[left=of img1,node distance=0cm,rotate=90,anchor=center,yshift=-1.1cm]{\large{Level spacing ratio $r$}};
\node[left=of img1,node distance=0cm,yshift=2.0cm,xshift=3cm]{\large{a.}};
\node (img2) [below=of img1,yshift=1.55cm]{\includegraphics[width=\linewidth]{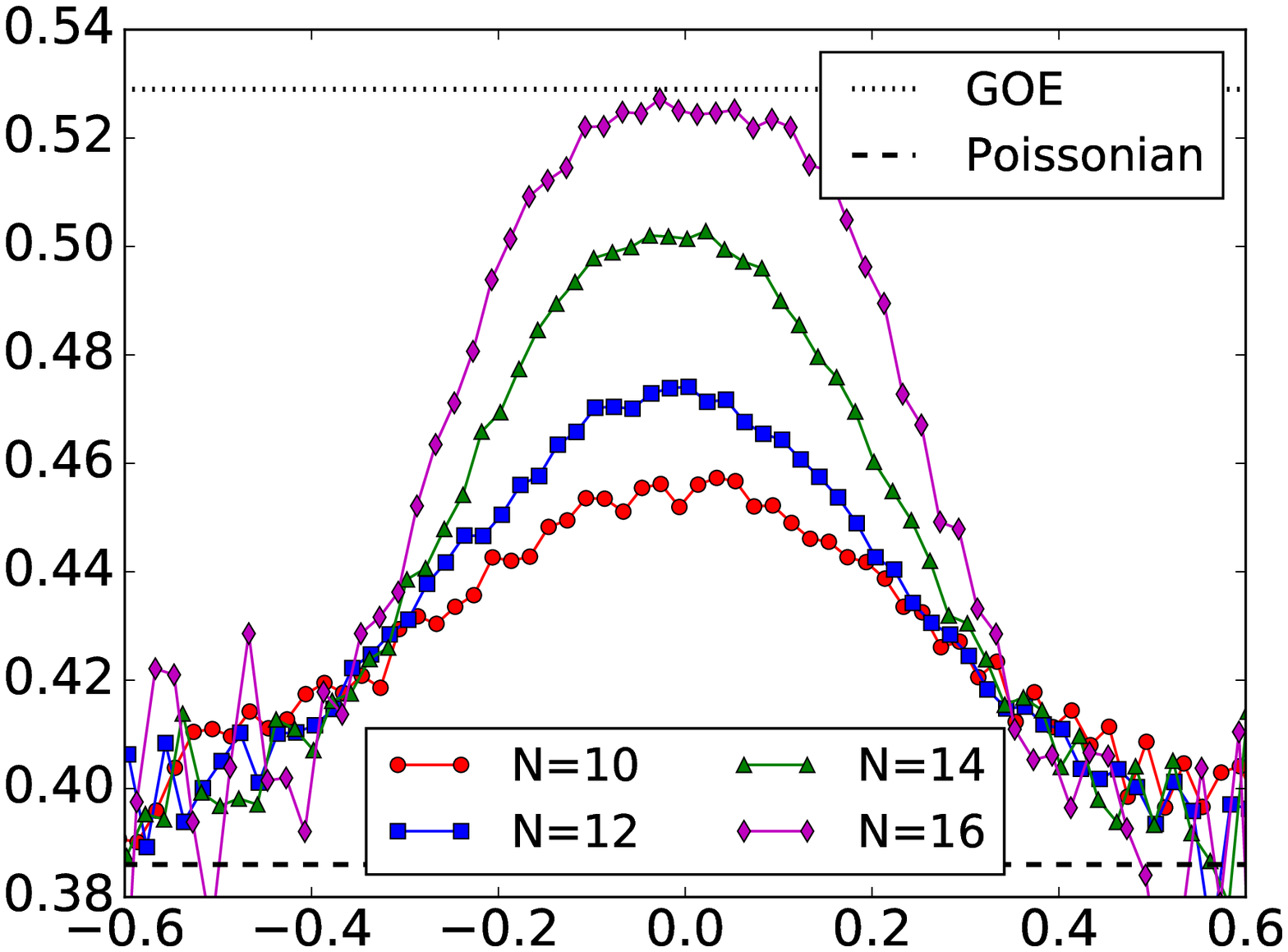}};
\node[left=of img2,node distance=0cm,rotate=90,anchor=center,yshift=-1.1cm]{\large{Level spacing ratio $r$}};
\node [below= of img2,node distance=0cm,yshift=1.5cm]{\large{Energy density $\varepsilon$}};
\node[left=of img2,node distance=0cm,yshift=2.0cm,xshift=3cm]{\large{b.}};
\end{tikzpicture}
\caption{Variation of level spacing ratio as a function of energy density for different exponents in Case-II: The level spacing ratio is plotted as a function of energy density for $p=2$[a.] and $p=5$[b.]. There are thermal states in the middle of the spectrum and localized states at the tail of the band. The energy window of mid spectrum thermal states decreases with increasing $p$.}
\label{fig:exp}
\end{figure}

For a better understanding, we have plotted the level spacing ratio as a function of energy density of the many body states for different values of the exponent. Fig.~\ref{fig:exp} shows that both for $p=2$ [a.] and $p=5$ [b.], there are the mid-spectrum states which follow the GOE distribution implying ergodicity and the states away from the middle of the spectrum follows Poissonian statistics indicating localization. There are mobility edges separating the two phases in both the cases. It also shows that for larger $p(=5)$, the energy window, for which the many body states are ergodic, is smaller than that in case of $p=2$. This causes the decrease in average level spacing ratio. However this window does not decrease to zero as $p$ is increased, instead it saturates to a value corresponding to the case-I.

We have shown (in appendix-B) that the covariance $Cov(r)=m\left(1-\frac{r}{N}\right)^p$ corresponds to a real space Hamiltonian which describes spinless fermions in a one dimensional system interacting through long ranged $p$-particle interactions with random interaction strengths. However, to be physically meaningful, the. allowed values of $p$ cannot be greater than the particle number $m$. Thus, in Fig.~\ref{fig:r_exp}, the average level spacing ratio vs. $p$ curves have to be truncated at $p=m=\frac{N}{2}$ for each value of $N$. Nevertheless, the basic inference drawn from the figure does not change with this truncation as the system is still ergodic for $p>1$.
Fig.~\ref{fig:r_exp} thus shows that with increasing $p$, which is the number of particles connected through random interactions in Eqn.~\ref{eq:random_ham}, the size of the energy window of mid-spectrum thermal states decreases and the tail of localized states increases before saturating for a large enough $p$ (which corresponds to case-I).

\paragraph{\textbf{\emph{Varying}} \boldmath{$\gamma$, $p=1:$}}
Here we have chosen $p=1$, i.e., the variation of the covariance with the Hamming distance is linear.
By changing the variable $\gamma$, we go from a limit of the same covariance for all pairs of distinct sites to the case where two sites a Hamming distance $N$ away are completely anti-correlated with the covariance decreasing linearly with Hamming distance.

\begin{figure}[t]
\begin{tikzpicture}
\node (img) {\includegraphics[width=\linewidth]{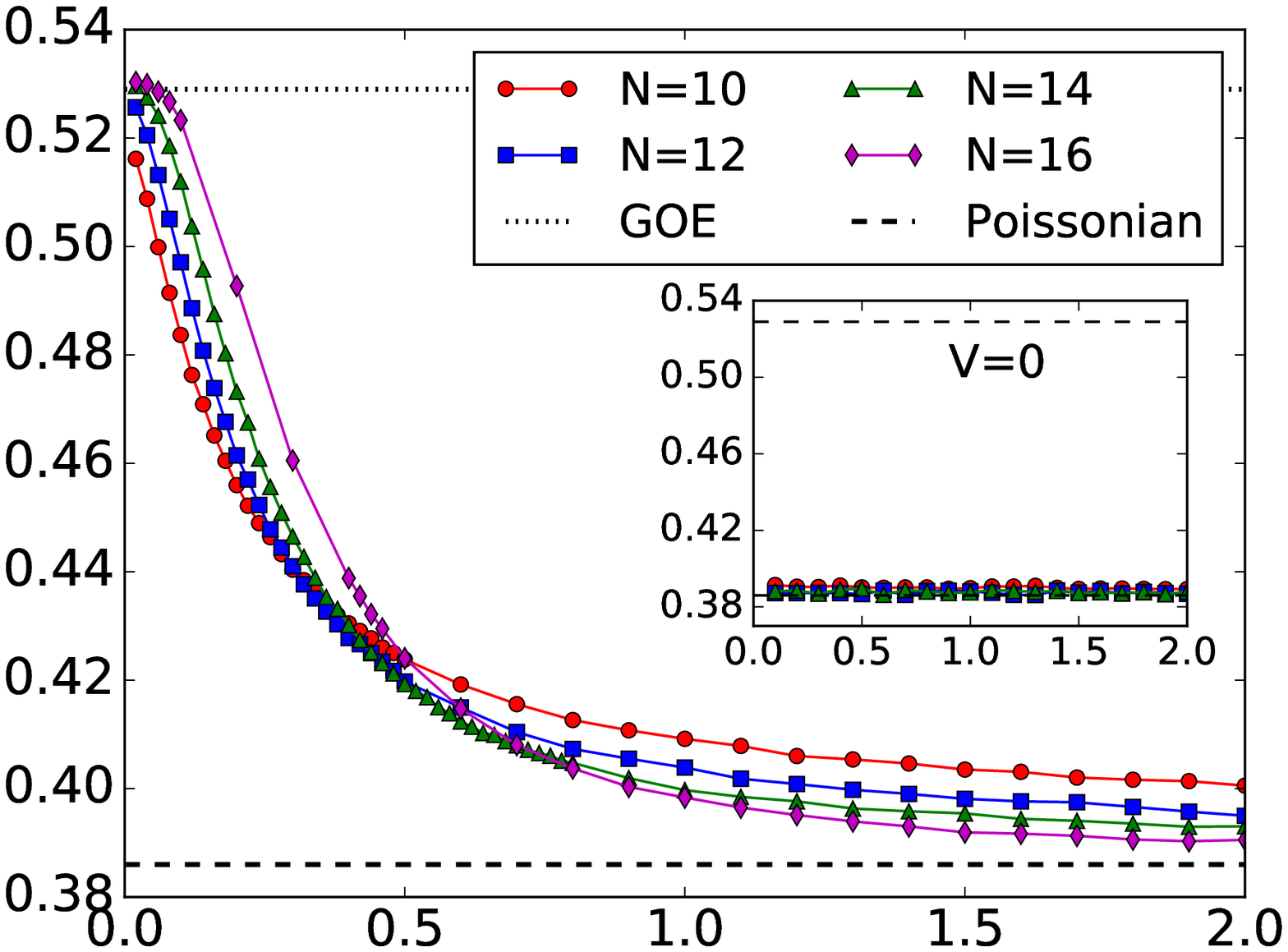}};
\node[below=of img,node distance=0cm,yshift=1.5cm]{\large{$\gamma$}};
\node[left=of img,node distance=0cm,rotate=90,anchor=center,yshift=-1.1cm]{\large{Average level spacing ratio $\braket{r}$}};
\end{tikzpicture}
\caption{Variation of the average level spacing ratio with $\gamma$ for $Cov(r)=m(1-\gamma\frac{r}{N})$: The average level spacing ratio is plotted as a function of $\gamma$ in the presence of interaction (and in the absence of interaction in inset). In the absence of interaction the average level spacing ratio remains at its Poissonian value for the whole range of $\gamma$. For non-zero interaction strength $V$, the average level spacing ratio changes from the GOE to Poissonian value implying an ergodic to MBL transition as $\gamma$ is changed from 0 to 2. From the crossover of finite size calculations the lower bound for the critical value of $\gamma$ for this transition can be defined. The interaction strength is chosen equal to the strength of the nearest neighbor hopping ($=0.2$) and the average level spacing ratio is calculated by averaging over $5000,2000,500,200$ samples for $L=10,12,14,16$ respectively.}
\label{fig:gamma}
\end{figure}

Fig.~\ref{fig:gamma}(inset) shows the variation of the average level spacing ratio of the system with changing $\gamma$.
For this linear variation of covariance with the Hamming distance, the many body states always obey Poissonian statistics. This can imply two things: 1) the system is always many body localized for the whole range of $\gamma$, or 2) the system is somehow integrable.

Now for two values of $\gamma$ ($\gamma=0$ and $1$), we know the corresponding real space Hamiltonians exactly.
For $\gamma=0$, all the disorder potentials in Fock space are completely correlated which implies the same on-site term on the entire Fock space.
This scenario corresponds to a real space Hamiltonian
\begin{equation}
H=\sum_i \left[-t_1\left( c_{i+1}^\dagger c_i+ h.c. \right) -t_2\left( c_{i+2}^\dagger c_i + h.c. \right)+\mu n_i\right]
\label{eq:t_t'_v_model}
\end{equation}
where $t_1,t_2$ and $\mu$ are the nearest neighbor hopping, next nearest neighbor hopping and chemical potential respectively. This is an integrable model. However upon introducing nearest-neighbor interactions, the integrability breaks down and the system thermalizes ~\cite{PRB.Subroto}. Additionally, for $\gamma=1$, the Hamiltonian in Fock space corresponds to a real space Hamiltonian describing an interacting Anderson insulator as described in appendix-A. Therefore we know that for $\gamma=1$ the system is in the MBL phase. So introducing a local interaction like the nearest neighbor interaction does not thermalize the system. It merely increases the effective strength of the interaction already present at $\gamma=1$.
These two facts motivate us to introduce a real space nearest neighbor interaction term $V\sum_i n_i n_{i+1}$ for all values of $\gamma$, where the interaction strength is chosen equal to the nearest neighbor hopping amplitude $t_1$.

Fig.~\ref{fig:gamma} shows the variation of the average level spacing ratio with $\gamma$ in the presence of the interaction $V$.
It can be clearly seen that the presence of the interaction not only thermalizes the system near $\gamma=0$ but also for a wider range of values of $\gamma$.
There is a thermal to MBL transition with increasing $\gamma$ and from the crossing of curves corresponding to different system sizes, one can define a lower bound of the critical value $\gamma_c$ below which the system is always thermal.
As expected this value of $\gamma_c$ depends on the strength of the interaction and shifts towards larger values for increasing interaction strength.
\paragraph*{\textbf{\emph{Case-III:}}}
In case-II(a.) we see that a change of the exponent $p$ from $p=1$ to $p=2$ abruptly changes the average level spacing ratio from the Poissonian value to GOE value. Here we interpolate between these two values of $p$. As changing the value of $p$ from $1$ to $2$ continuously is not allowed due to the breaking of positive semi-definiteness of the covariance matrix for non-integer exponents $p$, we interpolate between the two cases by mixing terms with exponents $1$ and $2$ as follows.
\begin{equation}
Cov(r)=m\left[(1-a)\left(1-\frac{r}{N}\right)+a\left(1-\frac{r}{N}\right)^2\right]
\end{equation}
When $a$ changes from $0$ to $1$, the covariance relation changes from the case $p=1$(for $a=0$) to the case $p=2$(for $a=1$) with an admixture of the two cases for intermediate values of $a$.
\begin{figure}[t]
\begin{tikzpicture}
\node (img) {\includegraphics[width=\linewidth]{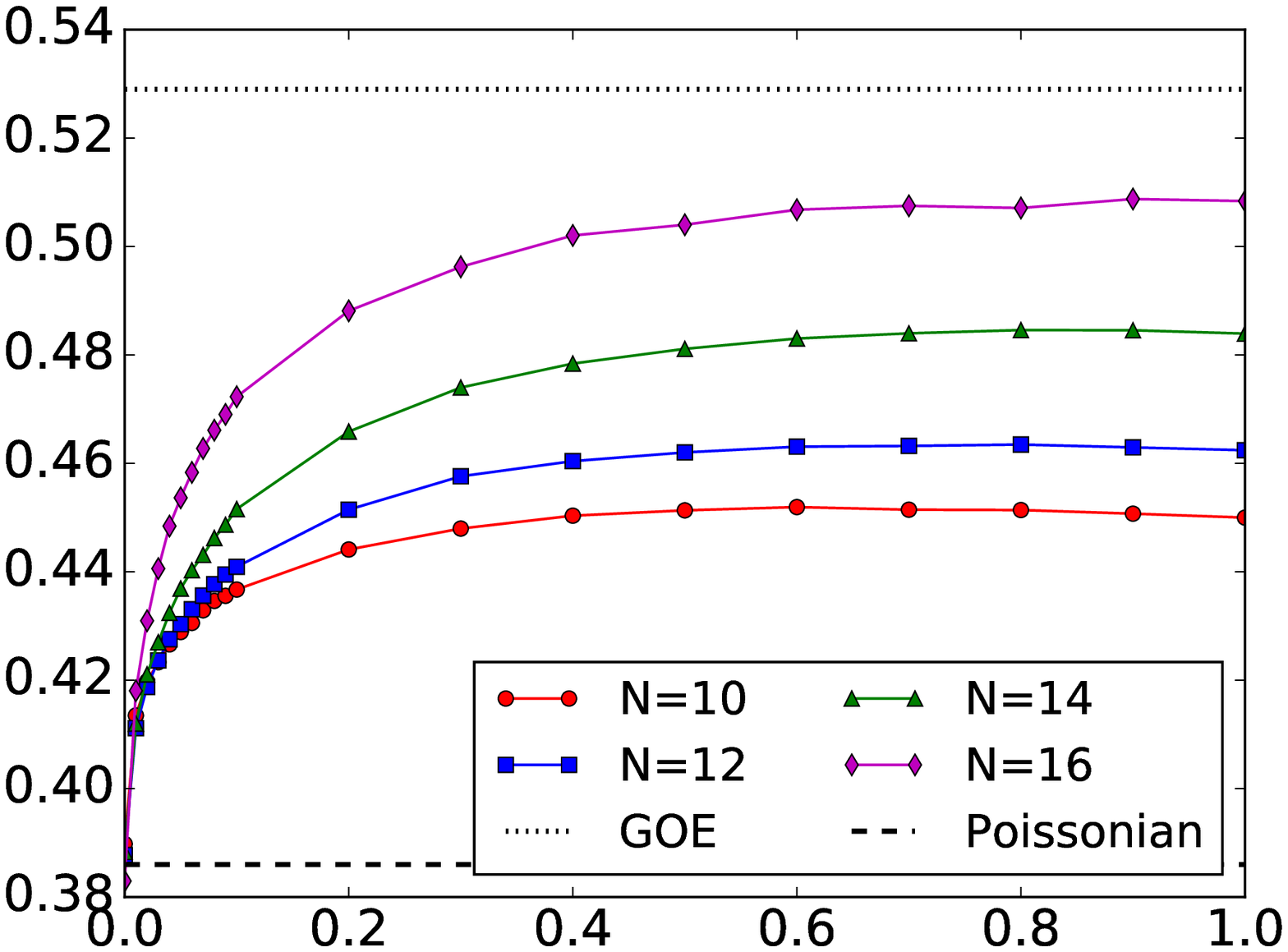}};
\node[below=of img,node distance=0cm,yshift=1.5cm]{\large{$a$}};
\node[left=of img,node distance=0cm,rotate=90,anchor=center,yshift=-1.1cm]{\large{Average level spacing ratio $\braket{r}$}};
\end{tikzpicture}
\caption{Variation of average level spacing ratio with $a$ for $Cov(r)=m\left[(1-a)(1-\frac{r}{N}) +a(1-\frac{r}{N})^2\right]$: It starts off at the Poissonian value for $a=0$ and increases abruptly towards the GOE value for very small non-zero value of $a$. This increase becomes sharper for larger system sizes indicating ergodic phase for non-zero value of $a$ in the thermodynamic limit. The average level spacing ratio is calculated by averaging over $5000,2000,500,50$ samples for $L=10,12,14,16$ respectively.}
\label{fig:exp-12}
\end{figure}

Fig.~\ref{fig:exp-12} shows the variation of average level spacing ratio as the parameter $a$ is changed from $0$ to $1$. For $a=0$, the average level spacing ratio has the Poissonian value which is expected as for this value the system is known to be in an MBL phase. However as $a$ is even slightly increased from $0$, the average level spacing ratio reaches a value near the GOE value indicating ergodicity. This shift becomes larger with increasing system sizes confirming ergodic behavior in thermodynamic limit. Once again there is no crossing of the finite size curves implying that there is no scale invariant critical value of the parameter $a_c$ which separates the MBL phase from thermal phase.

As discussed in the previous subsection, the covariance relation with exponent $p=2$ corresponds to a real space Hamiltonian representing random long-ranged two particle interactions.
Thus, the parameter $ma$ can be identified as the variance of these random interactions (see appendix-B). Therefore, Fig.~\ref{fig:exp-12} implies that as soon as random long-ranged interactions are introduced in a one dimensional system with random on-site disorder in real space, no matter how small the strength of the interaction is compared to the on-site disorder, the system always thermalizes.

\paragraph*{\textbf{\emph{Case-IV:}}}
\begin{figure}[t]
\begin{tikzpicture}
\node (img) {\includegraphics[width=\linewidth]{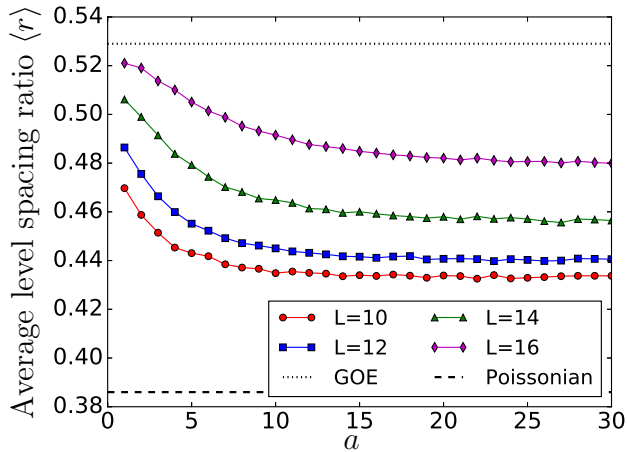}};
\node[below=of img,node distance=0cm,yshift=1.5cm]{\large{$a$}};
\node[left=of img,node distance=0cm,rotate=90,anchor=center,yshift=-1.1cm]{\large{Average level spacing ratio $\braket{r}$}};
\end{tikzpicture}
\caption{Variation of average level spacing ratio with $a$ for $Cov(r)=m~exp[-a\frac{r}{N}]$: For small values of $a$ the average level spacing ratio has a value near the GOE value and with increasing $a$ it decreases before finally saturating at values corresponding to case-I depending on the system size. The average level spacing ratio never reaches the Poissonian value implying absence of MBL phase. The average level spacing ratio is calculated by averaging over $5000,2000,500,50$ realizations of disorder for $L=10,12,14,16$ respectively.}
\label{fig:exponential}
\end{figure}
Here we calculate the average level spacing ratio of the system with the Hamiltonian of Eqn.~\ref{eq:ham1}, where the correlated on-site terms in Fock space have the covariance relation:
\begin{equation}
Cov(r)=m~exp\left[-a\dfrac{r}{N}\right]
\end{equation}
where $a$ is a parameter.
We vary this parameter in the range $[1,30]$.
The covariance here has an exponential decay with the Hamming distance $r$.
The decay of covariance becomes faster with increasing $a$ and for very large $a$ the covariance almost decays to zero before the Hamming distance reaches $r=2$.
In other words with, increasing $a$ the covariance relation becomes that of uncorrelated random potentials.
So, this model like the one in case-II(a.), approaches the model with uncorrelated Fock space disorder(case-I) in the large $a$ limit.

Fig.~\ref{fig:exponential} shows the variation of the average level spacing ratio with increasing $a$.
It is evident that for small $a$, where the covariance decays slowly with the Hamming distance, the average level spacing ratio has a value near the GOE value.
With increasing $a$ the average level spacing ratio decreases and saturates to a value which corresponds to the case-I. There is again no crossing between the curves corresponding to different sizes and the average level spacing ratio never comes close to the Poissonian value, which implies the absence of an ergodic to MBL phase transition. Instead, the system always remains thermal for the whole range of the parameter considered. The decrease in average level spacing ratio is due to the fact that for small $a$, almost all states are thermal while for larger $a$ only the mid-spectrum states are thermal with the states at the tail of the band being localized.

\paragraph*{\textbf{\emph{Case-V:}}}
Here, we calculate the average level spacing ratio of the system with Hamiltonian(Eqn.~\ref{eq:ham1}), where the correlated on-site terms in Fock space have the covariance relation:
\begin{equation}
Cov(r)=m\left[ u \delta_{r,0}+v\left(1-\delta_{r,0}\right)\right]
\end{equation}
We choose $u$ to be 1 and $v$ is varied from 0 to 1.
\begin{figure}[t]
\begin{tikzpicture}
\node (img) {\includegraphics[width=\linewidth]{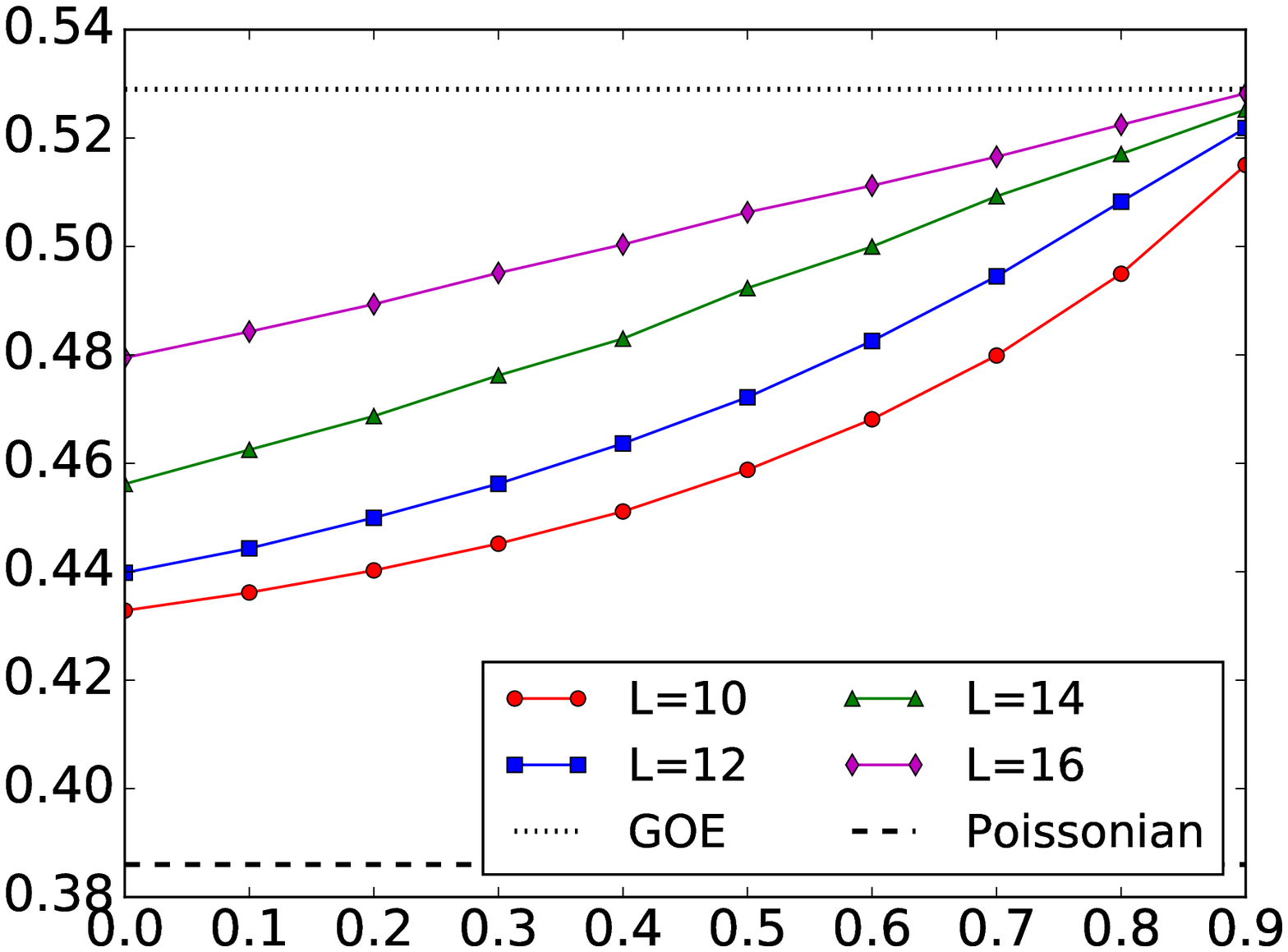}};
\node[below=of img,node distance=0cm,yshift=1.5cm]{\large{$v/u$}};
\node[left=of img,node distance=0cm,rotate=90,anchor=center,yshift=-1.1cm]{\large{Average level spacing ratio $\braket{r}$}};
\end{tikzpicture}
\caption{Variation of average level spacing ratio with $v/u$ for $Cov(r)=m~\left[u\delta_{r,0}+v\left(1-\delta_{r,0}\right)\right]$: For $v/u$ near zero, the average level spacing ratio has a value close to the value for case I and approaches the GOE value as $v/u$ approaches 1. The quantity $\braket{r}$ never reaches Poissonian value implying absence of MBL phase. The average level spacing ratio is calculated by averaging over $5000,2000,500,50$ realizations of disorder for $L=10,12,14,16$ respectively.
The level spacing ratio never reaches the Poissonian value and remains near the GOE value over the whole range of $v/u$. In this range the curves for different sizes never cross each other indicating the absence of a phase transition.}
\label{fig:uv}
\end{figure}
For $v=0$ the covariance matrix is a diagonal matrix with the diagonal terms equal to $m$.
This represents uncorrelated random numbers at Fock space sites with the variance of the random numbers being $m$(case-I).
On the other hand when $v=1$ all the on-site terms on Fock space sites are completely correlated, that is the on-site terms are all the same, which corresponds again to the Hamiltonian in Eqn.~\ref{eq:t_t'_v_model}.

Fig.~\ref{fig:uv} shows the average level spacing ratio with the variation of $v/u$.
For $v=0$ the average level spacing ratio starts from the value corresponding to case-I and approaches the GOE value for $v/u$ tending to $1$ as expected. There is no crossing between the curves for different system sizes and the average level spacing ratio never reaches the Poissonian value, which indicates absence of any ergodic-MBL phase transition.
\section{Conclusions}
We have investigated the effect of correlations between on-site disorder in Fock space on ergodicity and localization in one dimensional spinless fermionic systems. We have used covariance as a measure of correlations and Hamming distance on the Fock lattice as a measure of distance in Fock space and have examined the presence (or absence) of an ergodic-MBL transition for different kinds of functional forms of covariance with Hamming distance. Apart from the linear variation of covariance, all other cases display ergodicity in the system, with only a few localized many body states at the edge of the spectrum separated from ergodic states by a mobility edge in most cases.
Only in the case of a linear variation of the covariance with Hamming distance the level spacing statistics show Poissonian statistics implying non-thermal behavior in the system.
However that can imply either emergence of MBL phase which is believed to have emergent conservation laws~\cite{modak1.2015,Huse.Pheno.2014,Serbyn.2013,Chandran.2015} or conventional integrability of integrable system with dynamical symmetries that inhibit ergodicity~\cite{Integrable.Rigol,Chaos.Rigol,Integrable.Modak,RME.Modak}.
After the introduction of additional interactions which can break the integrability of the system, the system shows a thermal to MBL transition as a function of the slope of the linear variation of covariance (Fig.~\ref{fig:gamma}).

We also note that in case-I where all the Fock space disorder potentials are uncorrelated, the behavior of the system corresponds to that of QREM despite having a completely different structure of hopping in Fock space.
Thus, we conclude that the correlation between Fock space disorder affects ergodicity and localization in one dimensional systems. Strong enough disorder in Fock space with the correct form of covariance (linear variation) can drive thermal-MBL transition with decreasing correlation strength while with a different form of covariance the same strength of Fock space disorder does not give a MBL phase. Further, out of the set of different correlations we study only those with a linear variation with Hamming distance possess an ergodic-MBL transition.
The interacting Anderson model which is known to have an MBL phase for large enough disorder is a special case ($\gamma=1$ in our notation) of this type of linear variation.

\paragraph*{Acknowledgments:} SM thanks the UGC-ISF Indo-Israeli joint research program for funding.


%

\appendix
\section{Derivation of the covariance between on-site terms in Fock space for the interacting Anderson insulator}
Let us consider the Hamiltonian for the interacting Anderson model in a one dimensional system of spinless fermions:
\begin{equation}
H=\sum_i \varepsilon_i n_i -t\sum_i c_i^\dagger c_{i+1}+h.c.+V\sum_i n_i n_{i+1}
\label{eq:app-anderson}
\end{equation}
where $\varepsilon_i$'s are the random on-site potentials providing the randomness in the system, $t$ is the parameter for hopping between nearest neighbors and $V$ is the nearest neighbor interaction. It is known for strong enough disorder, this model has many body localized phase.~\cite{PRB.Huse.2007}

We assume the system has $N$ sites and $m$ particles and the $\varepsilon_i$'s are Gaussian distributed random numbers with mean $0$ and variance $\sigma^2$.
The `Fock space' sites are denoted by $\{n_i\}$, where $\{n_i\}$ is the set of occupation numbers of individual sites in real space, which means there are exactly $m$ $1$'s and $(N-m)$ $0$'s in each set $\{n_i\}$.
Therefore for any disorder realization, the on-site terms in Fock space corresponding to two different Fock space sites $\{n_i\}$ and $\{m_i\}$ are:

\begin{eqnarray}\nonumber
& &U_{\{n_i\}}=\sum_i \varepsilon_i n_i +V\sum_i n_i n_{i+1}\\ \nonumber
& &U_{\{m_i\}}=\sum_i \varepsilon_i m_i +V\sum_i m_i m_{i+1}
\end{eqnarray}

Note that the hopping parameter $t$ does not appear in these on-site terms as it connects different Fock space sites and appears only in the off-diagonal terms of the Hamiltonian written in `Fock space' basis.

Here, we derive the covariance between these on-site terms $U_{\{n_i\}}$ and $U_{\{m_i\}}$ assuming the sites $\{n_i\}$ and $\{m_i\}$ are a Hamming distance $r$ away from each other, (i.e. $n_i\neq m_i$ for exactly $r$ number of sites in real space). Thus,
\begin{equation}\label{eq:app-derive}
Cov(r)=\left\langle U_{\{n_i\}}U_{\{m_i\}} \right\rangle-\left\langle U_{\{n_i\}} \right\rangle \left\langle U_{\{m_i\}} \right\rangle
\end{equation}
where $\braket{\dots}$ implies an average over different realizations of the real space on-site disorder $\varepsilon_i$'s.
Throughout the derivation we will assume that the Gaussian random variables $\varepsilon_i$ are identically distributed and uncorrelated, so that
\begin{equation}\nonumber
\braket{\varepsilon_i \varepsilon_j}=\sigma^2\delta_{i,j}
\end{equation}
Therefore,
\begin{eqnarray}\nonumber
\left\langle U_{\{n_i\}}U_{\{m_i\}} \right\rangle &=&\left\langle \sum_{i,j}\varepsilon_i \varepsilon_j n_i m_j+V\sum_{i,j} \varepsilon_i n_i m_j m_{j+1}\right .\\ \nonumber
& &\left .+V\sum_{i,j} \varepsilon_j m_j n_i n_{i+1}\right .\\ \nonumber
& &\left .+V^2\sum_{i,j} n_i n_{i+1} m_j m_{j+1}\right\rangle\\ \nonumber
&=&\sigma^2\sum_{i,j} \delta_{i,j} n_i m_j +V^2 \sum_{i,j} n_i n_{i+1} m_j m_{j+1}
\end{eqnarray}
where we have used the fact $\braket{\varepsilon_i}=0$ to discard the two summations in the middle.
Similarly,
\begin{eqnarray}\nonumber
\left\langle U_{\{n_i\}} \right\rangle&=&\left\langle \sum_i \varepsilon_i n_i+V\sum_i n_i n_{i+1}\right\rangle\\ \nonumber
&=&V\sum_i n_i n_{i+1}
\end{eqnarray}
Therefore, Eqn.~\ref{eq:app-derive} becomes,
\begin{eqnarray}\nonumber
Cov(r)&=&\sigma^2\sum_{i,j} \delta_{i,j} n_i m_j\\ \nonumber
&=&\sigma^2 \sum_{i} n_i m_i
\end{eqnarray}

Now, $n_im_i$ is not equal to zero only when $n_i=m_i=1$. Since we are dealing with a constant number of particles only even hamming distances are allowed between any two pair of Fock space sites. If $r$ is the hamming distance between them then exactly $r/2$ real space sites become unoccupied while going from one Fock space site to another. Therefore the number of sites $i$ for which $n_i=m_i=1$ is $(m-\frac{r}{2})$.
Therefore,
\begin{eqnarray}
Cov(r)&=&\sigma^2\left(m-\frac{r}{2}\right)\\ \nonumber
&=&m\sigma^2\left(1-\frac{r}{N}\right)
\end{eqnarray}
where we use the fact that we are working at half filling.
\section{Derivation of the covariance between on-site terms in Fock space in case of a random $p$-particle interaction in real space}
Suppose we have a Hamiltonian of the from
\begin{equation}\label{eq:app-pham1}
H=\sum_{i_1,i_2,\dots,i_p}J_{i_1 i_2 \dots i_p} n_{i_1} n_{i_2}\dots n_{i_p}
\end{equation}
which represents spinless fermions in one dimension interacting through random long-ranged $p$-particle interactions given by $J_{i_1 i_2 \dots i_p}$s which are random numbers.
It is easy to extend the derivation given in Appendix-A to this case. Here, the covariance varies as a function of the Hamming distance as
\begin{equation}\label{eq:app-covp}
Cov(r)=m^p\sigma_p^2\left(1-\frac{r}{N}\right)^p
\end{equation}
where we assume  $J_{i_1 i_2 \dots i_p}$'s are Gaussian distributed random numbers with mean $0$ and variance $\sigma_p^2$.

However, the Hamiltonian in Eqn.~\ref{eq:app-pham1} is not very physical because if we have a one dimensional system with $N$ sites and $m$ particles interacting through a $p$-particle interaction there should be only one $J_{i_1 i_2 \dots i_p}$ corresponding to a particular set of sites $\{i_1,i_2,\dots,i_p\}$. Instead in Eqn.~\ref{eq:app-pham1} there are different $J_{i_1 i_2 \dots i_p}$'s for different permutations of ${i_1,i_2,\dots, i_p}$. Also note that $p$ cannot be greater than $m$, the particle number which is fixed beforehand. Nonetheless, Eqn.~\ref{eq:app-pham1} can be written in a form
\begin{eqnarray}\nonumber
H&=&\sum_{i_1<i_2<i_3<\dots<i_p}\left(J_{i_1 i_2 \dots i_p}+\dots\right)n_{i_1} n_{i_2}\dots n_{i_p}+\\ \nonumber
& &\sum_{i_1<i_2<\dots<i_{p-1}} \left(J_{i_1 i_1 i_2\dots i_p}+\dots\right)n_{i_1} n_{i_2} \dots n_{i_p}+\dots\\ \nonumber
& &+\sum_{i1<i2}\left(J_{i_1 i_1 \dots i_2 i_2}+\dots\right)n_{i_1} n_{i_2}+\sum_{i_1} J_{i_1 i_1 i_1\dots i_1} n_{i_1}
\end{eqnarray}
where the variables over which the summations are performed are restricted to reduce multiple counting of sets ${i_1,i_2,\dots,i_p}$. Clearly the expressions in the brackets are summations of different random numbers $J_{i_1 i_2 \dots i_p}$'s for different permutations of a particular set $\{i_1,i_2,\dots,i_p\}$ and will yield random numbers which have variances $\sigma_p^{'2}$ different from the initial variance of the $J$'s.
In other words the Hamiltonian(Eqn.~\ref{eq:app-pham1}) can be written in a form:
\begin{eqnarray}\label{eq:app-phamf}
H&=&\sum_{i_1<i_2<\dots<i_p} \alpha_{i_1 i_2 \dots i_p} n_{i_1} n_{i_2} \dots n_{i_p}+\\ \nonumber
& &\sum_{i_1<i_2<\dots<i_{p-1}} \beta_{i_1 i_2 \dots i_{p-1}} n_{i_1} n_{i_2} \dots n_{i_{p-1}}+\dots\\ \nonumber
& &+\sum_{i_1<i_2<i_3} \zeta_{i_1 i_2 i_3} n_{i_1} n_{i_2} n_{i_3}+\sum_{i_1<i_2} \eta_{i_1 i_2} n_{i_1} n_{i_2} +\sum_{i_1} \varepsilon_i n_i
\end{eqnarray}
where there is no multiple counting for a particular set of sites $\{i_1,i_2,\dots,i_p\}$ and $\alpha$,$\beta$,$\dots$,$\zeta$,$\eta$,$\varepsilon$ are new set of random numbers. If the variances of these random numbers are carefully chosen, it can be shown that Eqn.~\ref{eq:app-phamf} exactly corresponds to Eqn.~\ref{eq:app-pham1}.

For example let us consider the simplest case $p=2$.
Here the Hamiltonian(Eqn.~\ref{eq:app-pham1}) can be written as follows
\begin{eqnarray}\nonumber
H&=&\sum_{i,j} J_{ij}n_i n_j\\ \nonumber
&=&\sum_{i<j}(J_{ij}+J_{ji}) n_i n_j +\sum_i J_{ii}n_i\\ \nonumber
&=&\sum_{i<j}\eta_{ij} n_i n_j+\sum_i \varepsilon_i n_i
\end{eqnarray}
where $\eta_{ij}=(J_{ij}+J_{ji})$ are random numbers with variance $2\sigma_2^2$ and $\varepsilon_i=J_{ii}$ are random numbers with the variance $\sigma_2^2$ given $\sigma_2^2$ is the variance of the random numbers $J_{ij}$'s in Eqn.~\ref{eq:app-pham1}.

Similarly for $p=3$ the Hamiltonian (Eqn.\ref{eq:app-pham1}) is:
\begin{equation} \nonumber
\mathcal{H}=\sum_{i<j<k} \zeta_{ijk} n_i n_j n_k +\sum_{i<j} \eta_{i,j} n_i n_j +\sum_i \varepsilon_i n_i
\end{equation}
where $\zeta_{ijk}$, $\eta_{ij}$ and $\varepsilon_i$ are random numbers with variances $3!\sigma_3^2$, $6\sigma_3^2$ and $\sigma_3^2$ respectively.
Thus for any $p$, we avoid multiple counting and write the Hamiltonian in Eqn.~\ref{eq:app-pham1} in the form of Eqn.~\ref{eq:app-phamf} where there are single particle random on-site terms, two particle random interaction terms, three particle random interaction terms,$\dots$, $p$ particle random interaction terms for a one dimensional system of spinless fermions with $N$ sites. For all types of interactions, the random interaction strengths have variances which are some number times the variance $\sigma_p^2$.

We know from~\cite{REM.Derrida} that to get an extensive energy for the $p$-spin model, the $p$-spin random interaction term must have a variance which is inversely proportional to $N^{p-1}$ where $N$ is the size of the system. Similarly, here since we are dealing with $p$-particle interactions, the random numbers $J$'s in Eqn.~\ref{eq:app-pham1} should have a variance $\sigma_p^2 \propto\frac{1}{N^{p-1}}$. Consequently all the random interactions in Eqn.~\ref{eq:app-phamf} will have variances inversely proportional to $N^{p-1}$ since they are derived from $J$'s. Now, the $p$-particle interaction term with random interaction strength $\alpha_{i_1i_2\dots i_p}$ in Eqn.~\ref{eq:app-phamf} will give an extensive energy due to the choice of $\sigma_p^2$. However for the same choice the $(p-1)$ particle interaction term with random interaction strength $\beta_{i_1i_2\dots i_{p-1}}$ in Eqn.~\ref{eq:app-phamf} will contribute an energy independent of the system size and the corresponding term with $(p-2)$ particles interacting via a random interaction strength will give an energy contribution inversely proportional to system size $N$ {and so on}. Thus, in the thermodynamic limit only the $p$-particle interactions in Eqn.~\ref{eq:app-phamf} will contribute to the energy spectrum of the system, the others being negligibly small for a large system size $N$. Thus, we show that although the Hamiltonian in Eqn.~\ref{eq:app-pham1} seems unphysical, in the thermodynamic limit it is equivalent to a Hamiltonian
\begin{equation}\label{eq:random_ham}
\mathcal{H}=\sum_{i_1<i_2<\dots<i_p}J_{i_1 i_2\dots i_p} n_{i_1} n_{i_2} \dots n_{i_p}
\end{equation}
which represents particles in a one dimensional system interacting through $p$-particle interaction where the interaction strengths are Gaussian distributed random numbers with variance inversely proportional to $N^{p-1}$.
Therefore the covariance relation in Eqn.~\ref{eq:app-covp} also corresponds to Hamiltonian~\ref{eq:random_ham}  in the thermodynamic limit.
If we choose the variance $\sigma_p^2$ of random $J$'s to be $\left(\frac{2}{N}\right)^{p-1}$, which is possible since it satisfies the required scaling of the variance with system size for giving an extensive energy, we can write Eqn.~\ref{eq:app-covp} as
\begin{equation}
Cov(r)=m\left(1-\frac{r}{N}\right)^p
\end{equation}
which is nothing but the covariance relation in Case-II(a.) in the main text.
\section{Shannon Entropy}
Here we calculate the Shannon entropy which quantifies the spreading of the many body wave function in the Fock space lattice. For a many body wave function $\ket{\Psi}$, the Shannon entropy is defined by
\begin{equation}
S_2 = - \sum_{\{n_i\}}|\braket{\{n_i\}|\Psi}|^{2}log(|\braket{\{n_i\}|\Psi}|^{2})
\end{equation}
where the summation is over all the basis states  in Fock space lattice and $|\braket{\{n_i\}|\Psi}|^2$ is the amplitude of the many body wave function $\ket{\Psi}$ at Fock space site $\{n_i\}$.

For an extended state in the Fock space lattice, the amplitude of the wave function at each Fock space lattice site will be $\sim \frac{1}{\sqrt{V_H}}$ where $V_H$ is the dimension of the Fock space lattice which is $^NC_m$ in our case. Therefore the Shannon entropy for such an extended state will be $log(V_H)$. On the other hand for a localized state in Fock space the amplitude $|\braket{\{n_i\}|\Psi}|^2$ will be $\sim 1$ for a few Fock lattice sites $N_l$ over which the localized state is spread. Therefore the Shannon entropy will be proportional to $log(N_l)$ in case of localized sites in Fock space. Here we plot $S_2/log(V_H)$ which will be $\sim 1$ in case of extended states and will be negligibly small in case of localized states.
\begin{figure}[t]
\begin{tikzpicture}
\node (img1) {\includegraphics[width=\linewidth]{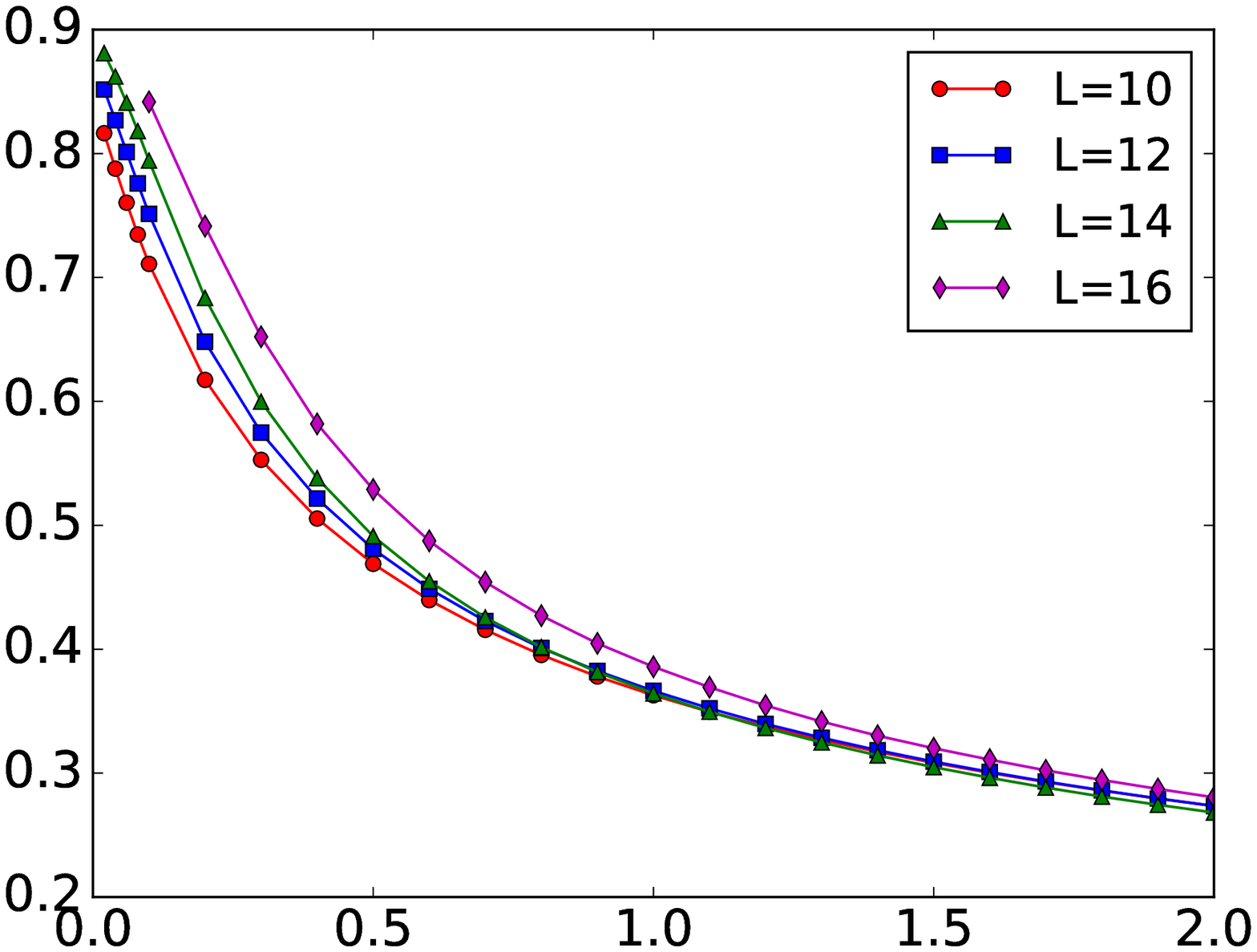}};
\node[left=of img1,node distance=0cm,rotate=90,anchor=center,yshift=-1.1cm]{\large{$S_2/log(V_H)$}};
\node[left=of img1,node distance=0cm,yshift=-1.5cm,xshift=3cm]{\large{a.}};
\node [below=of img1,node distance=0cm,yshift=6.0cm,xshift=-1cm]{\fbox{$V\neq 0$}};
\node (img2) [below=of img1,yshift=1.53cm]{\includegraphics[width=\linewidth]{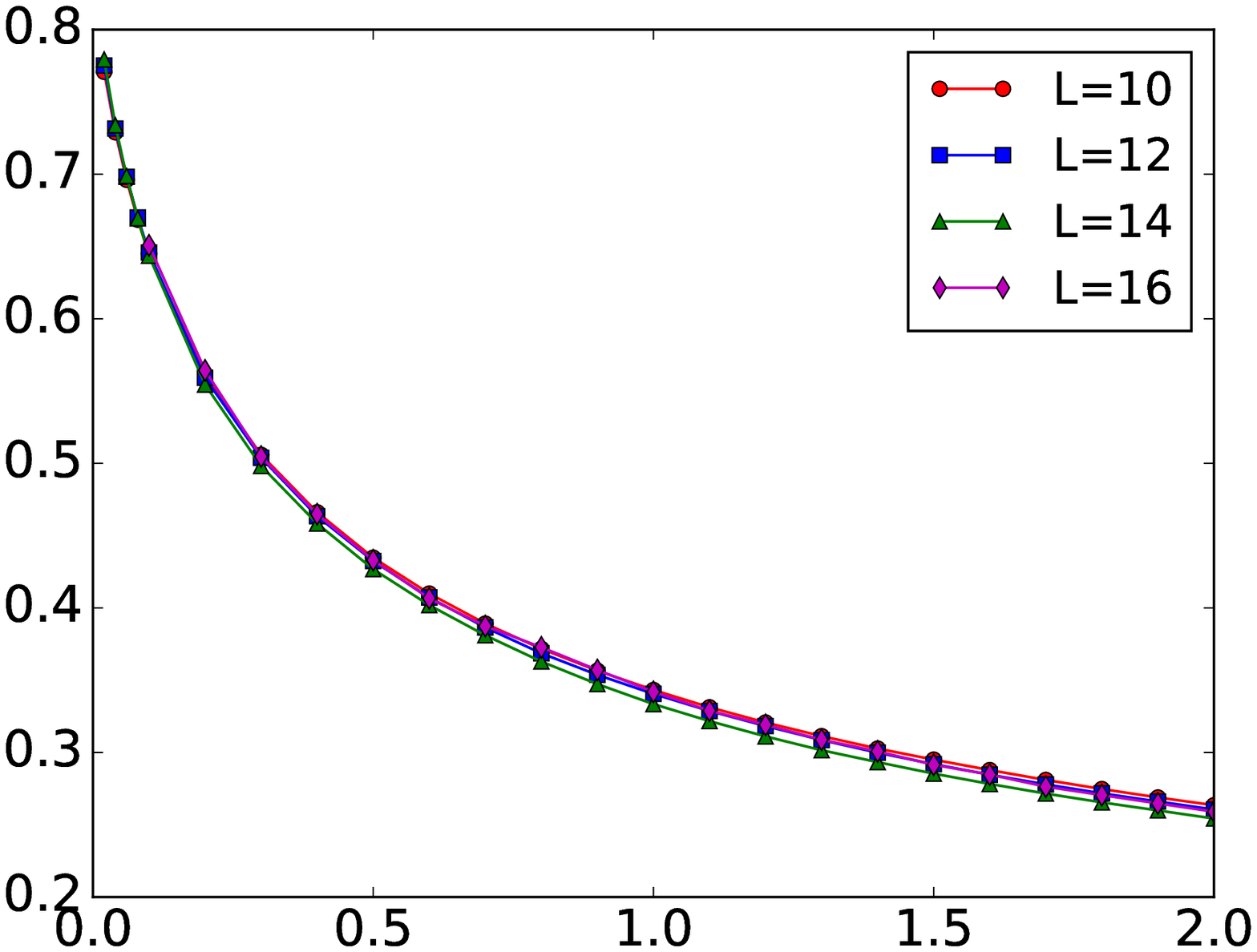}};
\node[left=of img2,node distance=0cm,rotate=90,anchor=center,yshift=-1.1cm]{\large{$S_2/log(V_H)$}};
\node [below= of img2,node distance=0cm,yshift=1.5cm]{\large{$\gamma$}};
\node [below= of img2,node distance=0cm,yshift=6.0cm,xshift=-1cm]{\fbox{$V=0$}};
\node[left=of img2,node distance=0cm,yshift=-1.5cm,xshift=3cm]{\large{b.}};
\end{tikzpicture}
\caption{Variation of Shannon entropy for functional form of covariance $cov(r)=m\left(1-\gamma \frac{r}{N} \right)$(case-II): The Shannon entropy in presence (a.) and absence (b.) of interaction has value $\sim log(V_H)$ at small values of $\gamma$ which indicates the extended behavior of many body wavefunctions. On the other hand,for large value of $\gamma$ it is small implying localization in Fock space.}
\label{fig:shannon}
\end{figure}

Fig.~\ref{fig:shannon} shows variation of $S_2/log(V_H)$ averaged over all the many body eigenstates for the Fock space Hamiltonian (Eqn.~\ref{eq:ham1}) with correlated on-site terms having covariance given by Eqn.~\ref{eq:case2}(Case-II) as a function of $\gamma$. Fig.~\ref{fig:shannon}(a) shows that in presence of interaction (see Section-IV Case-II) for small $\gamma$ the system has extended states in Fock space while for larger $\gamma$ the Shannon entropy goes as $\sim D log(V_H)$ where $D$ is a small number. This implies that the wavefunctions for larger values of $\gamma$ occupy smaller part of the whole Fock space, but still possess some scale which depends on the system size $N$. Thus the states at larger values of $\gamma$ can be called `localized' in the sense that it do not extend over the whole Fock space, but are different from true localized states where there would not be any system size dependence in Shannon entropy. The Shannon entropy does not show any crossing of the curves corresponding to different system sizes as the level spacing ratio does(Fig.~\ref{fig:gamma}). Therefore Shannon entropy is not a useful quantity to show crossover between extended and localized phase in Fock space.

Fig.~\ref{fig:shannon}(b) shows that in absence of interaction the Shannon entropy plot looks qualitatively similar to that in presence of interaction($V\neq 0$). Again for small $\gamma$ the system has extended states in Fock space and for large $\gamma$ the system has `localized' states in Fock space. This is in contrary to the behavior of level spacing statistics which shows Poissonian statistics for all values of $\gamma$ in absence of interaction(Fig.~\ref{fig:gamma}) because extended states are expected to be thermalizing.

However this is understandable from the fact that Shannon entropy is basis dependent and a wave function which is localized in some basis can look completely extended if the basis chosen for calculating Shannon entropy is not the basis where the wave function is peaked around some basis state. For example Bloch states in real space looks completely extended, but in momentum space it corresponds to a particular momentum. Thus the level spacing statistics, being independent of basis choice, determines the level repulsion (or absence of it) between the eigenstates and captures the ergodic (or non-ergodic) behavior of the system correctly which Shannon entropy does not.
\section{Entanglement entropy}
\begin{figure}[t]
\begin{tikzpicture}
\node (img) {\includegraphics[width=\linewidth]{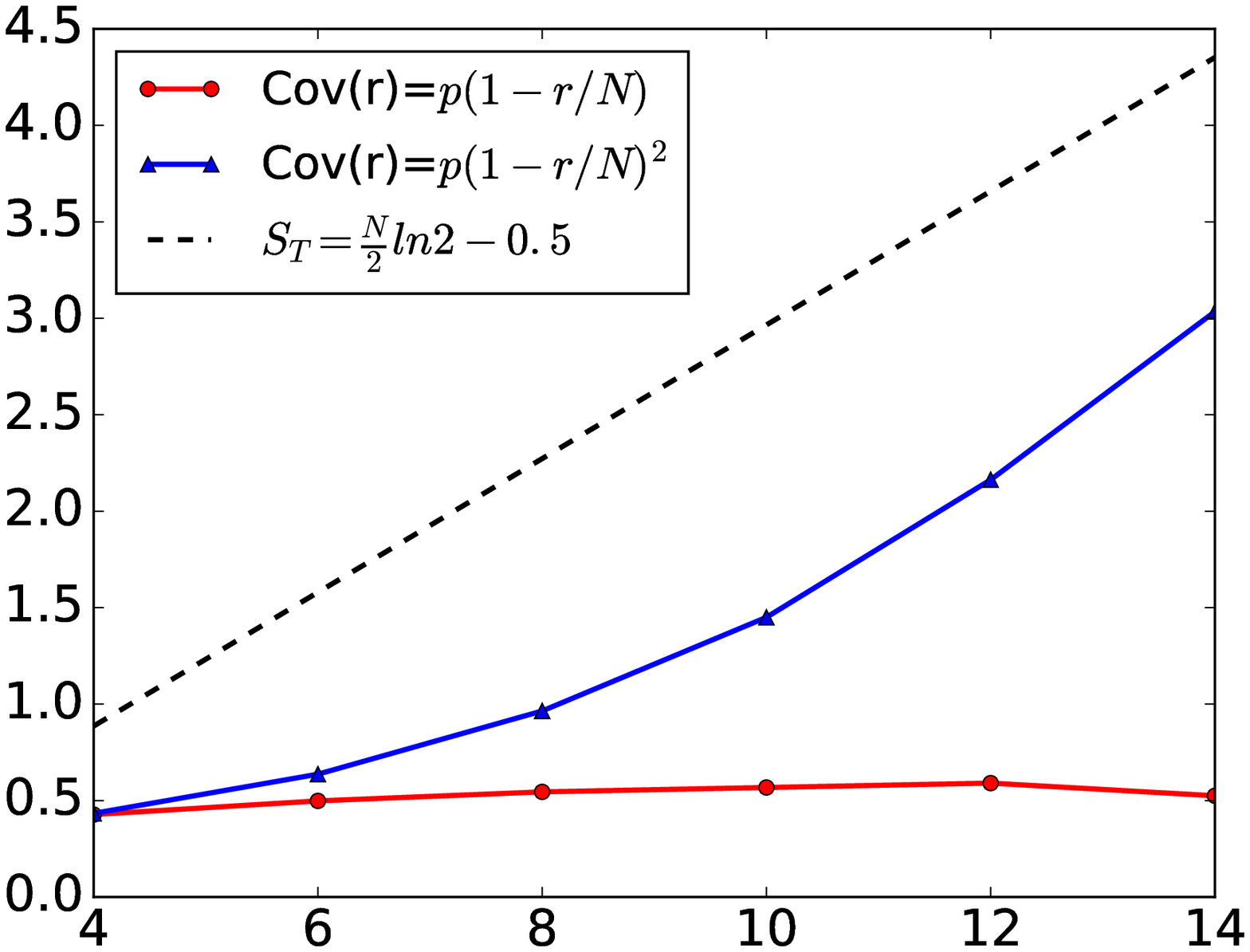}};
\node [below =of img,node distance=0,yshift=1.5cm]{\large{System size $N$}};
\node [left= of img,node distance=0,rotate=90,anchor=center,yshift=-1.1cm]{\large{Entanglement entropy}};
\end{tikzpicture}
\caption{The variation of the bipartite entanglement entropy as a function of system size: Bipartite entanglement entropy for one half of the system is plotted as a function of the total system size for $p=1$ and $p=2$ in case of covariance varying as Eqn.\ref{eq:case2}($\gamma=1$). The entanglement entropy is calculated by averaging over all eigenstates and a sufficiently large number of disorder realizations. The dashed line shows the variation of thermal entropy with the system size $N$.}
\label{fig:app-ee}
\end{figure}
The bipartite entanglement entropy can also be used to distinguish between ergodic and MBL phases.
It quantifies the correlation of quantum information between two parts of a system.
Here we calculate bipartite entanglement entropy between two halves of the system.
For typical eigenstates the entanglement entropy varies proportionately with the volume of the sub-system in the ergodic phase and is proportional to area of the sub-system in the MBL phase.
We calculate the Von Neumann entropy given by $S=-tr_B(\rho_Blog\rho_B)$ where $A$ and $B$ are the two halves of the system.
$\rho_B$ is the reduced density matrix obtained from the density matrix $\rho=\ket{\Psi}\bra{\Psi}$ by tracing out the degrees of freedom of sub-system $A$ where $\ket{\Psi}$ is a typical mid spectrum state of the many body system.
For one dimensional system of spinless fermions with $N$ lattice sites in real space the entanglement entropy $S$ becomes $\sim \frac{N}{2}$ in thermal phase and $\sim N^0$ in MBL phase.

Fig.~\ref{fig:app-ee} shows the scaling of entanglement entropy for $p=1$ and $p=2$ in case of covariance varying as Eqn.~\ref{eq:case2}.
It clearly shows that for $p=1$ the bipartite entanglement entropy scales as an area law($\sim N^0$) implying localization in the system while for $p=2$ it follows a volume law($\sim N/2$) indicating ergodicity.

\end{document}